\documentclass[%
aip,
amsmath,amssymb,
reprint,%
]{revtex4-1}
\usepackage[dvipsnames]{xcolor}
\usepackage[
	pdffitwindow=true,
	colorlinks=true,
	frenchlinks=false,
 linkcolor=blue,
	anchorcolor=blue,
 citecolor=blue,
 filecolor=blue,
 urlcolor=blue,
 bookmarks=true,
 bookmarksopen=true,
	bookmarksnumbered=true,
 bookmarksopenlevel=1,
 plainpages=false,
	pdfpagelayout=TwoPageLeft,
 pdfpagelabels=true,
	breaklinks
]{hyperref}
\usepackage[per-mode=symbol,separate-uncertainty]{siunitx}
\usepackage{graphicx}
\usepackage{dcolumn}
\usepackage{color}
\usepackage{graphicx,wrapfig,lipsum}
\usepackage{bm}
\usepackage[english]{babel}
\usepackage{color}
\usepackage{ulem}
\usepackage{float}
\usepackage{chemformula}

\begin{document}
%\preprint{AIP/123-QED}
\title[]{Growth optimization of TaN for superconducting spintronics}

\author{M.~M\"uller}
\email[]{manuel.mueller@wmi.badw.de}
\affiliation{Walther-Mei{\ss}ner-Institut, Bayerische Akademie der Wissenschaften, 85748 Garching, Germany}
\affiliation{Physik-Department, Technische Universit\"{a}t M\"{u}nchen, 85748 Garching, Germany}

\author{R.~Hoepfl}
\affiliation{Walther-Mei{\ss}ner-Institut, Bayerische Akademie der Wissenschaften, 85748 Garching, Germany}
\affiliation{Physik-Department, Technische Universit\"{a}t M\"{u}nchen, 85748 Garching, Germany}

\author{L.~Liensberger}
\affiliation{Walther-Mei{\ss}ner-Institut, Bayerische Akademie der Wissenschaften, 85748 Garching, Germany}
\affiliation{Physik-Department, Technische Universit\"{a}t M\"{u}nchen, 85748 Garching, Germany}

%\author{S.~Gepr{\"a}gs}
%\affiliation{Walther-Mei{\ss}ner-Institut, Bayerische Akademie der Wissenschaften, 85748 Garching, Germany}
\author{S.~Gepr\"ags}
\affiliation{Walther-Mei{\ss}ner-Institut, Bayerische Akademie der Wissenschaften, 85748 Garching, Germany}

\author{H.~Huebl}
\affiliation{Walther-Mei{\ss}ner-Institut, Bayerische Akademie der Wissenschaften, 85748 Garching, Germany}
\affiliation{Physik-Department, Technische Universit\"{a}t M\"{u}nchen, 85748 Garching, Germany}
\affiliation{Munich Center for Quantum Science and Technology (MCQST), Schellingstra{\ss}e 4, 80799 M\"{u}nchen, Germany}
\author{M.~Weiler}
\affiliation{Fachbereich Physik and Landesforschungszentrum OPTIMAS,
	Technische Universit\"{a}t Kaiserslautern, 67663 Kaiserslautern, Germany}
\affiliation{Walther-Mei{\ss}ner-Institut, Bayerische Akademie der Wissenschaften, 85748 Garching, Germany}
\affiliation{Physik-Department, Technische Universit\"{a}t M\"{u}nchen, 85748 Garching, Germany}
\author{R.~Gross}
\affiliation{Walther-Mei{\ss}ner-Institut, Bayerische Akademie der Wissenschaften, 85748 Garching, Germany}
\affiliation{Physik-Department, Technische Universit\"{a}t M\"{u}nchen, 85748 Garching, Germany}
\affiliation{Munich Center for Quantum Science and Technology (MCQST), Schellingstra{\ss}e 4, 80799 M\"{u}nchen, Germany}
\author{M.~Althammer}
\email[]{matthias.althammer@wmi.badw.de}
\affiliation{Walther-Mei{\ss}ner-Institut, Bayerische Akademie der Wissenschaften, 85748 Garching, Germany}
\affiliation{Physik-Department, Technische Universit\"{a}t M\"{u}nchen, 85748 Garching, Germany}
\date{\today}
\pacs{}
\keywords{}
\begin{abstract}
We have optimized the growth of superconducting TaN thin films on \ch{SiO2} substrates via dc magnetron sputtering and extract a maximum superconducting transition temperature of $T_{\mathrm{c}}=5$ K as well as a maximum critical field $\mu_0H_{\mathrm{c2}}=(13.8\pm0.1)$ T. To investigate the impact of spin-orbit interaction in superconductor/ferromagnet heterostructures, we then analyze the magnetization dynamics of both normal state and superconducting TaN/\ch{Ni80Fe20}(Permalloy, Py)-bilayers as a function of temperature using broadband ferromagnetic resonance (bbFMR) spectroscopy. The phase sensitive detection of the microwave transmission signal is used to quantitatively extract the inverse current-induced torques of the bilayers. The results are compared to our previous study on NbN/Py-bilayers. In the normal state of TaN, we detect a positive damping-like current-induced torque $\sigma_{\mathrm{d}}$ from the inverse spin Hall effect (iSHE) and a small field-like torque $\sigma_{\mathrm{f}}$ attributed to the inverse Rashba-Edelstein effect (iREE) at the TaN/Py-interface. In the superconducting state of TaN, we detect a negative $\sigma_{\mathrm{d}}$ attributed to the quasiparticle mediated inverse spin Hall effect (QMiSHE) and the unexpected manifestation of a large positive field-like $\sigma_{\mathrm{f}}$ of unknown origin matching our previous results for NbN/Py-bilayers. 
\end{abstract}
\maketitle
\section{Introduction}
Superconducting spintronics is a rapidly growing research field. An important aspect is the investigation of methods for injecting angular momentum into superconducting materials to benefit from the potentially enhanced spin transport properties in superconductors \cite{Birge2018, Linder2015, Blamire2014,Eschrig2015a, Ohnishi2020a}. First promising results include the detection of extremely long spin lifetimes \cite{ Yang2010} and large spin Hall angles (SHA) \cite{Wakamura2014, Wakamura2015} in superconductors as well as the detection of enhanced spin transport properties in$\;$superconductor(SC)/ferromagnet(FM)-bilayers \cite{Jeon2018}. An emerging direction within this young field is the investigation of superconductors with large spin-orbit interaction (SOI) \cite{Jacobsen2015,Bergeret2014, Nadj-Perge2014, Sajadi2018,Fatemi2018, Zhang2019} or in proximity to heavy metals \cite{Jeon2019b, Jeon2018}. This approach allows to study how SOI affects the superconducting and spin transport properties. Besides the generation of spin-triplet supercurrents \cite{Jacobsen2015,Bergeret2014}, theoretical predictions include the generation of supercurrents by Rashba-spin-orbit interaction \cite{Bychkov1984a, He2019}, the generation of Majorana quasiparticles \cite{Sato2012,Sato2010, Desjardins2019} and supercurrent-induced spin-orbit torques \cite{Hals2016}. For the fabrication of high quality superconducting spintronics devices, cheap, easy to fabricate and non-hazardous materials with large spin-orbit interaction are desirable. In this paper, we discuss TaN as a potential candidate for spin-current experiments with superconductors, motivated by the large SOI induced by Ta \cite{Chi2015,Liu2012}. We present the growth optimization of superconducting TaN deposited via reactive dc magnetron sputtering, a material which otherwise finds application in long-wavelength single photon detectors \cite{IlIn2012, Engel2012}, corrosion-resistive coatings \cite{Alishahi2016, Baba1996} and as a diffusion barrier against Cu \cite{Han2016, Cho1999, Laurila2002}. To investigate the impact of spin-orbit interaction on the magnetization dynamics parameters, we perform broadband ferromagnetic resonance (bbFMR) on TaN/\ch{Ni80Fe20}(Permalloy, Py)-bilayers in the normal conducting (NC) and superconducting (SC) state. The bbFMR experiments and in particular the frequency-dependent phase sensitive analysis of the transmission data allows to determine quantities such as the effective magnetization, magnetic anisotropies, and magnetization damping \cite{Muller, Berger2018}. In addition, this data also contains information about the electrical ac currents which arise from magnetization dynamics and can be attributed to inverse spin-orbit torques (iSOTs) as well as classical electrodynamics (i.e. Faraday's law) \cite{Berger2018, Meinert2020, Muller}. Both contributions can be quantified in terms of the complex inverse spin-orbit torque conductivity $\sigma^{\mathrm{SOT}}$, which relates the induced charge current in the sample to a change in magnetization $\boldsymbol{\mathrm{J}}\propto\sigma^{\mathrm{SOT}}\partial \boldsymbol{\mathrm{M}}/\partial t$ \cite{Freimuth2015}. Hence, bbFMR allows to simultaneously quantify the impact of an adjacent superconducting film in superconductor/ferromagnetic metal(FM)-heterostructures on both the magnetization dynamics (e.g. FMR linewidth) in the FM and on the spin-orbit torque conductivity $\sigma^\mathrm{SOT}=\sigma_\mathrm{f}+i\cdot\sigma_\mathrm{d}$, which comprises damping-like torques such as the inverse spin Hall effect (iSHE) in $\sigma_\mathrm{d}$ and field-like effects such as the inverse Rashba-Edelstein effect (iREE) \cite{Bychkov1984a, Edelstein1990} and Faraday currents in $\sigma_\mathrm{f}$. Our analysis is based on Ref. \onlinecite{Berger2018} and was adapted for SC/FM-heterostructures in our previous work \cite{Muller}. Comparing our results on TaN/Py bilayers to those of Ref. \onlinecite{Muller} reveals that SOI in TaN has a strong effect on the $\sigma^{\mathrm{SOT}}$ in the SC state. We in particular observe a more strongly pronounced $\sigma_{\mathrm{d}}$ in the superconducting state of our TaN/Py-bilayer. This further supports our conjecture in Ref. \ \onlinecite{Muller}, that a negative $\sigma_{\mathrm{d}}$ in the SC state originates from the quasiparticle mediated inverse spin Hall effect (QMiSHE) \cite{Wakamura2015}. \\
\section{Experimental Details}
For this study, we grow 60 nm thick TaN thin films on a thermally oxidized Si (100) substrate by reactive dc magnetron sputtering in a mixed Ar/\ch{N2}-atmosphere. For the SC/FM-heterostructures,
Py was deposited in situ on TaN to prevent oxidation at the SC/FM interface. Furthermore, a protecting $\mathrm{TaO_x}$ cap layer has been used resulting in TaN/Py/$\mathrm{TaO_x}$ heterostructures. 
% The layer thickness of TaN ($d_{\mathrm{SC}}=60$ nm) was chosen to be sufficiently thick to ensure an experimentally accessible $T_{\mathrm{c}}$. The layer thickness of Py ($d_{\mathrm{FM}}=5$ nm) was selected to be thick enough to ensure a reasonable signal-to-noise ratio and sufficiently thin to detect possible changes in the Gilbert damping $\alpha$ from altered spin-pumping properties in the SC state.
The thickness $d_{\mathrm{TaO_x}}$ of the $\mathrm{TaO_x}$ cap layer is chosen thin enough to have no measurable effect on the iSOT conductivities ($d_{\mathrm{TaO_x}}<2$ nm). Atomic force microscopy measurements on both TaN thin films and TaN/Py-bilayers reveal a surface roughness below detection limit ($<$300 pm) (see Fig. S1 of the Supplementary Material (SM) \onlinecite{Supplements}). For the determination of the resistivity and the superconducting transition temperature $T_{\mathrm{c}}$, we performed 4-point transport measurements
using the Van-der-Pauw-method \cite{VanderPAUW1991a} on single layer TaN films as a function of temperature between 3 K and 300 K. The exemplary data set in Fig. \ref{Fig: series1}(a) shows the resistivity during the transition from the normal to the superconducting state. 
 The gray dashed vertical line defines $T_{\mathrm{c}}=(T_1+T_2)/2$ as the mean value between the temperatures where $\rho$ takes on 90\% ($\rho(T_1)/\rho_{\mathrm{NC}}$=90\%) and 10\% ($\rho(T_2)/\rho_{\mathrm{NC}}$=10\%) of its normal state resistivity $\rho_{\mathrm{NC}}$ at $T=7$ K. This particular film was grown at a gas flow ratio \ch{N2}/Ar=0.35, a deposition temperature $T_{\mathrm{depo}}=500^\circ$C, a deposition pressure $p_{\mathrm{depo}}=5\;\mu$bar and a sputtering power $P=30$ W, resulting in $T_{\mathrm{c}}=4.97$ K. The area highlighted in blue marks the superconducting transition width $\Delta T_{\mathrm{c}}=T_1-T_2=1.26$ K. \\
Additionally, we performed x-ray reflectometry measurements (see Fig. \ref{Fig: series1}(b)) to determine the layer thickness $d_{\mathrm{SC}}$ and volume density $D_{\mathrm{V}}$ of our samples. For this particular film, we find $d_{\mathrm{SC}} \approx64$ nm and $D_{\mathrm{V}}\approx 13.8\;\mathrm{\;g/cm^3}$ in agreement with the expected density for cubic TaN \cite{Alishahi2016}.
In our experiments we used films with a thickness ranging between 54 and 66 nm.\\ 
\section{Results and Discussion}
\subsection{Growth optimization of superconducting TaN}
\begin{figure}[htpb]
	\centering
	\includegraphics[scale=1.0]{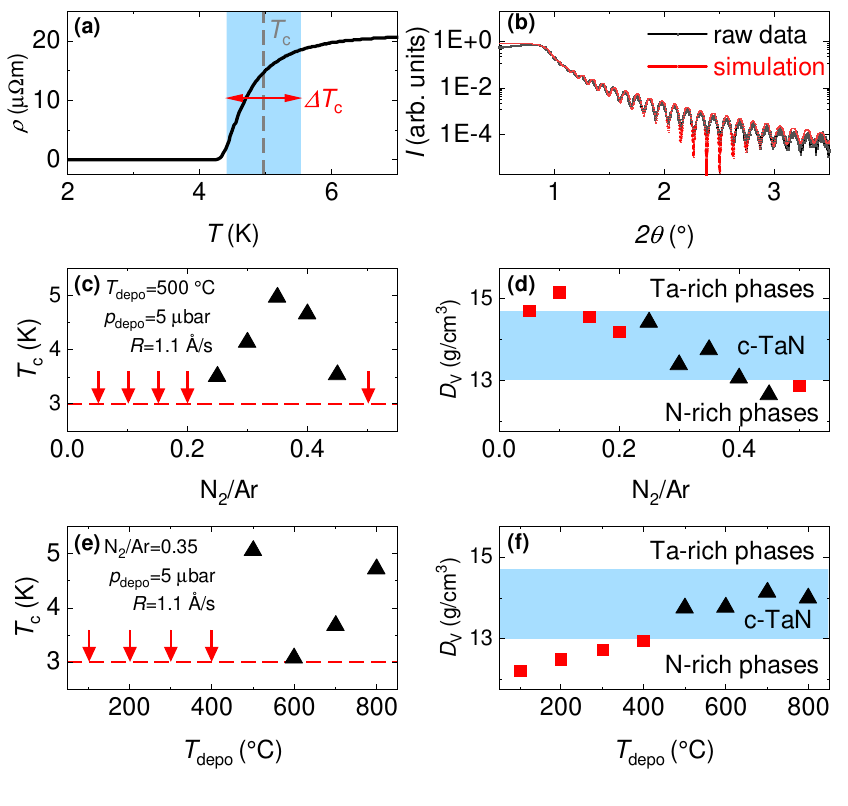}
	\caption{(a) Exemplary plot of the measured resistivity $\rho$ versus temperature $T$ together with the extracted superconducting transition temperature $T_{\mathrm{c}}$ (vertical dashed line) and transition width $\Delta T_{\mathrm{c}}$ (blue area). (b) Exemplary x-ray reflectometry data (black) and simulation curve (red) allowing to extract the layer thickness $d_{\mathrm{TaN}}$ as well as the volume density $D_{\mathrm{V}}$ of the TaN thin films. (c), (d) Superconducting transition temperature $T_{\mathrm{c}}$ and volume density $D_{\mathrm{V}}$ of the TaN thin films as a function of the $\mathrm{\mathrm{N_2/Ar}}$-flow ratio during deposition. (e), (f) $T_{\mathrm{c}}$ and $D_{\mathrm{V}}$ as a function of the deposition temperature $T_{\mathrm{depo}}$. Samples that did not exhibit full superconductivity above $T=3$ K are represented by red arrows in (c) and (e) as well as red squares in (d) and (f).
	}
	\label{Fig: series1}
\end{figure}
In order to optimize their superconducting properties, we deposited TaN-films at a deposition temperature of $T_{\mathrm{depo}}=500^\circ$C and a pressure of $p=5\mathrm{\;\mu}$bar at varying $\mathrm{N_2/Ar}$ flow ratios during deposition. Figure \ref{Fig: series1}(c) shows the transition temperature $T_{\mathrm{c}}$ as function of the \ch{N2}/Ar-flow ratio during the deposition. For flow ratios $(0.25\leq \mathrm{N_2/Ar}\leq 0.45)$, we find a $T_{\mathrm{c}}$ of the TaN thin films above 3 K. In addition, we mark gas flow mixtures that did not exhibit full superconductivity above $T=3$ K by red arrows. For this optimization series, we find a maximum of $T_{\mathrm{c}}=4.97$ K at a gas flow ratio of $\mathrm{N_2/Ar}=0.35$. Notably, this ratio is significantly higher compared to earlier studies \cite{Shin2001, Wakasugi1997, IlIn2012, IlIn2012a}, which predict a maximal $T_{\mathrm{c}}$ of TaN at gas flow ratios \ch{N2}/Ar = 0.10-0.15. This difference can have various reasons such as a different distance between target and substrate or different values of the kinetic energy of the Ta-atoms during the sputtering process due to varying sputtering power $P$ and target bias voltage (see Fig. S2 of the SM \cite{Supplements}). \\
From the extracted volume density $D_{\mathrm{V}}$ in Fig. \ref{Fig: series1}(d), we can identify three regimes with different TaN phases in agreement with the results in Ref. \onlinecite{Alishahi2016}. In particular, we find non-superconducting regimes for low and high \ch{N2}-concentrations, which we attribute to the formation of Ta- and N-rich $\mathrm{TaN}_\mathrm{{1\pm\delta}}$ phases such as \ch{Ta_2N} and \ch{TaN_2} for low and high N2/Ar-ratios, respectively. The intermediate range $(13.0\leq D_{\mathrm{V}}\leq 14.7) \mathrm{\;g/cm^3}$ is attributed to the cubic TaN-phase (c-TaN) \cite{Alishahi2016}. In the fabricated films, all three phases may be present in parallel, with varying volume fractions, since in our experiments we only can determine the average volume density. For the optimized samples with the highest $T_{\mathrm{c}}$, however, we expect a high fraction of our thin films to be in the c-TaN phase. In Fig. \ref{Fig: series1}(e), we plot the $T_{\mathrm{c}}$ of our thin films as a function of the deposition temperature $T_{\mathrm{depo}}$ for $p = 5\;\mu$bar and \ch{N2}/Ar=0.35. It becomes apparent that superconducting films with a $T_\mathrm{c}$ higher than 3 K were found only for $T_{\mathrm{depo}}>400^{\circ}$C. Figure \ref{Fig: series1}(f) suggests that this threshold in temperature coincides with the transition of our samples from N-rich $\mathrm{TaN_{1+\delta}}$ to the c-TaN phase. Consequently, the optimized growth parameters for superconducting TaN with a high $T_{\mathrm{c}}$ are \ch{N2}/Ar=0.35, $T_{\mathrm{depo}}=500^\circ$C , $p_{\mathrm{depo}}=5\;\mu$bar and $P_{\mathrm{depo}}=30$ W.
Our maximum $T_\mathrm{c}$ values are smaller than those reported in Ref. \onlinecite{Chaudhuri2013}, where $T_{\mathrm{c}}=6$ K was achieved for c-TaN films grown on \ch{SiO2} using infrared pulsed laser deposition. TaN is known to have even higher superconducting transition temperatures of up to $T_{\mathrm{c}}=10.2$ K on other substrates such as \ch{Al2O3} (sapphire) \cite{IlIn2012, IlIn2012a} and \ch{MgO} \cite{Chaudhuri2013, Shin2001}, where films with a better texture can be achieved. However, our optimal $T_{\mathrm{c}}$ values for TaN are in agreement with previous results,indicating that thermally oxidized \ch{Si} is not an ideal substrate for the growth of TaN with maximum $T_\mathrm{c}$ values \cite{Chaudhuri2013}. To confirm this suggestion, we utilized the optimized growth parameters to deposit TaN films on c-plane sapphire substrates and obtained $T_{\mathrm{c}}$=7.72 K for a 60 nm thick TaN film. For these films we also found reflections from the c-TaN-phase by x-ray diffraction \cite{Chaudhuri2013, Hashizume2011} (see Fig. S3 of the SM \cite{Supplements}). We note, however, that the purpose of this study is not to maximize $T_{\mathrm{c}}$ but to compare the spin-orbit torques arising from superconducting TaN to those that we previously reported for NbN \cite{Muller} also grown on \ch{SiO2}.\\
\subsection{Critical magnetic field of optimized TaN films}
\begin{figure}[htpb]
	\centering
	\includegraphics[scale=1.0]{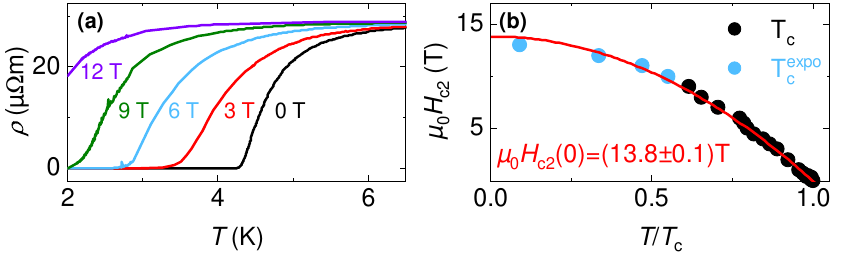}
	\caption{(a) Resistivity $\rho$ as a function of temperature $T$ measured in an external in-plane magnetic field $\mu_0H_{\mathrm{ext}}$ for a TaN film fabricated using the optimized parameters. (b) Extracted critical field $\mu_0H_{\mathrm{c2}}(T)$ (black and blue dots) as a function of reduced temperature $T/T_{\mathrm{c}}$ together with the fitted behavior of $\mu_0H_{\mathrm{c2}}(T)$ following Eq. (\ref{Eq: Critical-Field}). The blue data points indicate the linearly extrapolated $T_{\mathrm{c}}$ for high external magnetic fields, where the SC thin film did not exhibit full superconductivity for the lowest experimentally achievable temperature.
	}
	\label{Fig: series2}
\end{figure}

To analyze the magnetic field dependence of the superconducting properties of our TaN-films, we measured resistivity vs. temperature curves of our optimized TaN thin film (\ch{N2}/Ar=0.35, $T_{\mathrm{depo}}=500^\circ$C, $p_{\mathrm{depo}}=5\;\mu$bar and $P_{\mathrm{depo}}=30$ W, $d=60$ nm), for fixed in-plane magnetic fields in the range between 0 and 12 T. The result is shown in Fig. \ref{Fig: series2}(a). We observe a reduction in $T_{\mathrm{c}}$ as well as an increase of the superconducting transition width with increasing external magnetic field strength $\mu_0H_{\mathrm{ext}}$.
In Fig. \ref{Fig: series2}(b) we plot the transition temperature $T_{\mathrm{c}}(\mu_0H_{\mathrm{ext}})/T_{\mathrm{c}}(0)$ versus the applied in-plane field $\mu_0H_{\mathrm{ext}}$. Here, the $T_{\mathrm{c}}$ values ranging below the lowest experimentally accessible temperatures are determined by extrapolation. In order to estimate the upper critical field $H_{\mathrm{c2}}$ we fit the data to the empirical dependence
\begin{equation}
\mu_0H_{\mathrm{c2}}(T)=\mu_0H_{\mathrm{c2}}(T=0)\left[1-\left(\frac{T}{T_{\mathrm{c}}}\right)^{2}\right]
\label{Eq: Critical-Field}
\end{equation}
and find $\mu_0H_{\mathrm{c2}}(T=0)=(13.8\pm0.1)$ T, which agrees well to the value of $\mu_0H_{\mathrm{c2}}=14$ T reported in Ref. \onlinecite{IlIn2012} for TaN films grown on sapphire substrates. This result shows that our TaN films are sufficiently resilient to external magnetic fields in the 1 T-range, required for bbFMR experiments.\\
\subsection{BbFMR experiments and inductive analysis on TaN/Py-bilayers}
To investigate the spin transport properties of TaN in the normal and superconducting state, we study SC/FM bilayers, where the FM is a 5 nm thick \ch{Ni80Fe20} (Permalloy, Py) thin film. In the following we present results on two bilayers: (i) sample A is a TaN/Py bilayer, where the TaN layer is grown using the optimal deposition parameters (\ch{N2}/Ar=0.35, $T_{\mathrm{depo}}=500^\circ$C, $p_{\mathrm{depo}}=5\;\mu$bar, $P_{\mathrm{depo}}=30$ W and $d_{\mathrm{TaN}}=60$ nm, resulting in $T_{\mathrm{c}}=4.7$ K). (ii) sample B is a TaN/Py bilayer where the growth parameters of TaN (\ch{N2}/Ar=0.1, $T_{\mathrm{depo}}=500^\circ$C, $p_{\mathrm{depo}}=5\;\mu$bar, $P_{\mathrm{depo}}=30$ W and $d_{\mathrm{TaN}}=60$ nm) are chosen such that $T_{\mathrm{c}}<2.25$ K. The difference in the growth parameters is the lower \ch{N2}/Ar gas flow ratio which favors the formation of Ta-rich phases. This is supported by x-ray reflectometry and the extracted $D_\mathrm{V}$ values shown in Fig. S4 of the SM \cite{Supplements}. Sample B can therefore viewed as a reference sample allowing us to verify whether the experimentally observed signatures can be assigned to the formation of a superconducting phase in the TaN.
\begin{figure}[htpb]
	\centering
	\includegraphics[scale=0.249]{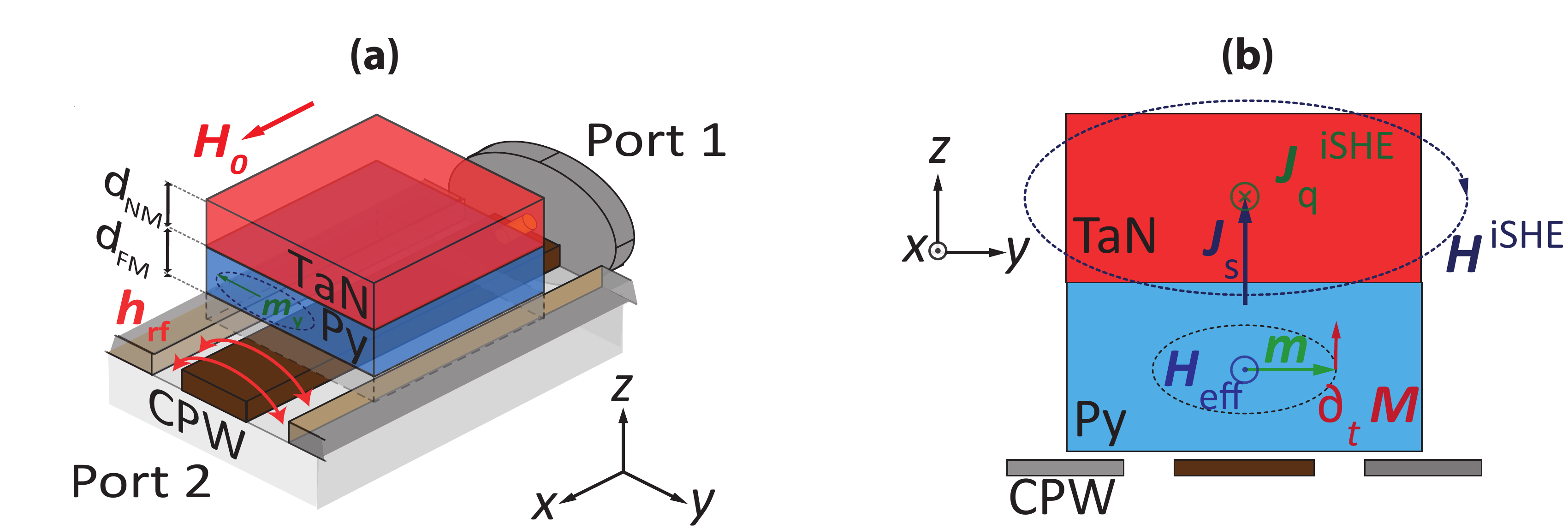}
	\caption{(a) Sketch of the experimental bbFMR setup including the measurement geometry for in-plane bbFMR. (b) Schematic illustration for the generation of the charge current density $\boldsymbol{\mathrm{J}}_\mathrm{q}^{\mathrm{iSHE}}$ by the ac iSHE. The ac flux $\boldsymbol{\mathrm{H}}^{\mathrm{iSHE}}$ generated by $\boldsymbol{\mathrm{J}}_\mathrm{q}^{\mathrm{iSHE}}$ is coupled into the coplanar waveguide (CPW).
	}
	\label{Fig: series3}
\end{figure}
 %in-situ deposited 5 nm of the ferromagnetic metal on top of two TaN films: One grown under the ptimized deposition conditions (sample A, ) and one under deposition parameters, which lead to normal state properties of the TaN film for all experimentally accessible $T$ (sample B). For sample B we chose identical growth properties to sample A except for a lower $\mathrm{N_2/Ar}$-ratio ($\mathrm{N_2/Ar}$=10\%). 
To access the magnetic properties and the spin current phenomena, we perform bbFMR measurements in a cryogenic environment over a broad temperature range. A sketch of the measurement setup and geometry is shown in Fig. \ref{Fig: series3}(a). The bilayers are mounted face-down onto a coplanar waveguide (CPW) and we record the complex microwave transmission parameter $S_{21}$ using a vector network analyzer (VNA). In particular, we measure the transmission $S_{21}$ for fixed microwave frequencies $f$ in the range ($5 \mathrm{\;GHz}\leq f\leq 35 \mathrm{\;GHz}$) as a function of the external magnetic field $H_{\mathrm{ext}}$, applied along the $\boldsymbol{\mathrm{x}}$-direction. We use a sufficiently small VNA microwave output power of 1 mW \cite{Bailleul2003, Neudecker2006}, such that all spin dynamics are in the linear regime. For $T>T_{\mathrm{c}}$, the magnetization dynamics excited in the Py pumps a spin current density $\boldsymbol{\mathrm{J}}_\mathrm{s}$ into the adjacent TaN layer as illustrated in Fig \ref{Fig: series3}(b). In the TaN layer, $\boldsymbol{\mathrm{J}}_\mathrm{s}$ is absorbed and converted into a charge current $\boldsymbol{\mathrm{J}}_\mathrm{q}^{\mathrm{iSHE}}$ via the inverse spin Hall effect (iSHE). On the one hand, the spin pumping effect manifests itself as an additional contribution to the Gilbert damping $\alpha$ of the FMR as it represents an additional relaxation channel for angular momentum \cite{Tserkovnyak2002}. On the other hand, the ac magnetic field $H^{\mathrm{iSHE}}$ generated via the iSHE induced ac charge current $\boldsymbol{\mathrm{J}}_\mathrm{q}^{\mathrm{iSHE}}$ is inductively coupled to the CPW and is detected by the VNA \cite{Berger2018}. All experiments are performed 
with $\boldsymbol{\mathrm{H}}_{\mathrm{ ext}}\parallel\boldsymbol{\mathrm{x}}$ to suppress the formation of superconducting vortices.\\
We extract the magnetization dynamics parameters by studying the resonance field $\mu_0H_{\mathrm{res}}$ and linewidth $\mu_0\Delta H$ as a function of excitation frequency $f$. The resonance field $\mu_0H_{\mathrm{res}}(f)$ is fitted to the ip-Kittel equation (Eq. (S1) in the SM \cite{Supplements}) to extract the effective magnetization $\mu_0M_{\mathrm{eff}}$, $g$-factor and in-plane anisotropy field $\mu_0H_{\mathrm{ani}}$. The frequency evolution of the FMR linewidth $\mu_0\Delta H(f)$ is fitted with the linear model to extract the Gilbert damping parameter $\alpha$ and the inhomogeneous broadening $\mu_0H_{\mathrm{inh}}$ \cite{Muller}. In this work, we are primarily interested in the temperature dependence of the Gilbert damping $\alpha(T)$ as it is sensitive to spin pumping across the interface $\alpha=\alpha_0+\alpha_{\mathrm{SP}}$, where $\alpha_0$ is the intrinsic damping and $\alpha_{\mathrm{SP}}$ denotes the additional dissipation channel for angular momentum via spin pumping \cite{Flacke2019, Suraj2020}.
For completeness, we present the temperature dependence of the remaining characteristic parameters describing the FMR in section V of the SM \cite{Supplements}.
\begin{figure}[htpb]
	\centering
	\includegraphics[scale=1.0]{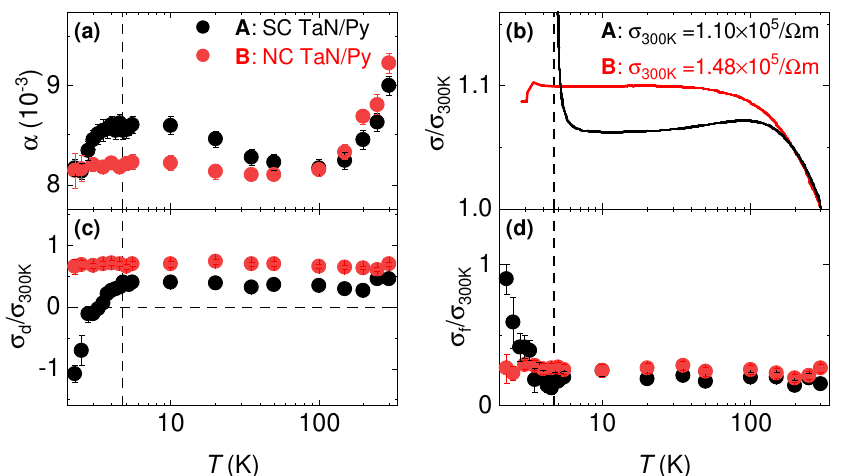}
	\caption{(a) Temperature dependence of the Gilbert damping parameter $\alpha$ for both the SC (black, A) and NC (red, B) TaN/Py-bilayer. The apparent decrease of $\alpha$ in the SC state is due to the suppression of spin pumping into the SC due to the freeze-out of thermally excited quasiparticles. (b) Normalized conductivity $\sigma/\sigma_{\mathrm{300K}}$ of both samples as a function of $T$. A small valley in $\sigma/\sigma_{\mathrm{300K}}$ for sample A in the range $(T_{\mathrm{c}}\leq T\leq 50 \mathrm{\;K})$ mirrors the increase in $\alpha$. (c) Damping-like iSOT conductivity $\sigma_{\mathrm{d}}$ normalized by the sample conductance at $T=300$ K $\sigma_{\mathrm{300K}}$ as a function of $T$. A positive $\sigma_{\mathrm{d}}$ is detected in both samples at high temperatures. In the SC state, the $\sigma_{\mathrm{d}}$ of sample A decays to negative values. (d) Field-like iSOT conductivity $\sigma_{\mathrm{f}}$ as a function of $T$. We observe a small positive $\sigma_{\mathrm{f}}$ for both samples in the normal state attributed to the iREE. In the SC state, $\sigma_{\mathrm{f}}$ rises to large positive values. These results are in good agreement to previous results on NbN/Py-bilayers \cite{Muller}.
			}
	\label{Fig: series4}
\end{figure}
The extracted Gilbert damping $\alpha$ in Fig. \ref{Fig: series4}(a) is very similar for both samples and, moreover, is in agreement to a previous study on Py films capped with a thin insulating TaN-layer \cite{Zhao2016}. The observed reduction in $\alpha(T)$ with decreasing $T$ down to $T=100$ K in both samples is compatible with previous results obtained for Py thin films \cite{Frangou2017, Suraj2020, Muller}. For sample A, we observe an increase in $\alpha$ from 100 K to 5 K, while TaN is in the NC state. This feature is mirrored in the normalized conductivity $\sigma/\sigma_{\mathrm{300 K}}$ of sample A, which was measured via dc resistance experiments and is plotted as a function of $T$ in Fig. \ref{Fig: series4}(b). In sample B, both $\alpha(T)$ and $\sigma(T)/\sigma_{\mathrm{300 K}}$ remain roughly constant in this temperature range. We hence interpret the increase in $\alpha(T)$ in sample A to an additional resistivity-like damping contribution of this TaN/Py-bilayer \cite{Heinrich1979,Liu2009}. In the SC state, the $\alpha$ of sample A decreases towards lower temperatures at first gradually and then at $T=2.5$ K abruptly drops to a low fixed value of $\alpha=8.2\cdot 10^{-3}$. In comparison, the $\alpha$ value of sample B remains roughly constant in this temperature range. The temperature dependence of $\alpha$ for sample A below $T_{\mathrm{c}}$ is attributed to the blocking of spin currents at the TaN/Py-interface, which is in agreement with previous results \cite{Muller, Bell2008}. For temperatures slightly below $T_{\mathrm{c}}$, thermally excited quasiparticles in the SC can still mediate spin currents. Hence, we observe a gradual reduction of $\alpha$ with decreasing $T$. The higher $\sigma_{\mathrm{300 K}}$ in Fig. \ref{Fig: series4}(b) for sample B containing TaN grown at a lower $\mathrm{N_2/Ar}$-ratio agrees with previous results \cite{Shin2001, Hashizume2011, Wakasugi1997, Nie2001}, and is most likely caused by the higher concentration of Ta-rich phases in the thin film.\\
By measuring both the FMR amplitude $A$ and phase $\phi$, we are able simultaneously extract the inverse spin-orbit torque conductivities $\sigma^{\mathrm{SOT}}$ with the data analysis procedure established in Ref. \onlinecite{Muller}. Exemplary raw data for the normalized inductive coupling $\tilde{L}$ between sample and CPW, which is the fundamental quantity that relates the FMR amplitude $A$ and phase $\phi$ to the $\sigma^{\mathrm{SOT}}$, is shown in the SM section VI \cite{Supplements}. In Fig. \ref{Fig: series4}(c), we plot the extracted damping-like iSOT $\sigma_{\mathrm{d}}$ for both samples normalized by $\sigma_{\mathrm{300K}}$. We observe a sizable positive $\sigma_{\mathrm{d}}$ for both samples in the normal state that is about temperature independent. This contribution comprises the combined iSHE effects of Permalloy and TaN in our samples. We have previously observed similar positive $\sigma_{\mathrm{d}}$ for NbN/Py-bilayers despite the negative spin Hall angle (SHA) in NbN \cite{Rogdakis2019, Wakamura2015}. Hence, we assume that the contribution of Py to $\sigma_{\mathrm{d}}$ dominates due to a higher $\sigma$ of Py and we can not quantify the magnitude of the SHA of TaN. However, we can state that the various TaN phases exhibit different SHA as evident from the larger $\sigma_{\mathrm{d}}$ of sample B containing primarily Ta-rich phases like \ch{Ta2N}. In the superconducting regime of sample A, $\sigma_{\mathrm{d}}$ first decreases towards zero with decreasing temperature due to a superconducting shunting effect of the TaN layer as previously detected in Ref. \cite{Muller}. Reducing the temperature further, it takes increasing negative values in agreement with the quasiparticle mediated inverse spin Hall effect (QMiSHE)\cite{Wakamura2014, Muller} assuming a negative SHA of TaN, like pure Ta \cite{Chi2015,Liu2012}. We attribute the abrupt manifestation of negative $\sigma_{\mathrm{d}}$ for $T\leq2.5 $ K to a second superconducting transition of the fraction of the TaN layer that is in close proximity to the Py layer. This is supported by analyzing the net inductive coupling between FM and CPW in the presence of Meißner currents in the SM section VII\cite{Supplements}. In previous studies, it has been shown, that proximity to a FM material can have a significant effect on the superconducting $T_\mathrm{c}$ \cite{Buzdin2005, Jiang1995, Khusainov1997}. For $(2.5\mathrm{\; K}<T<T_{\mathrm{c}})$, we thus assume that sample A does not exhibit a direct SC/FM-contact and therefore only observe the SC shunting effect of the iSHE. For lower $T$, the entire TaN layer becomes superconducting and the established direct SC/FM-contact enables the SC quasiparticles to contribute to the spin transport via the QMiSHE. It is important to note that in the SC state a diverging spin Hall angle resulting from intrinsic and side-jump contributions to the SHE compensates the diminishing quasiparticle population with decreasing $T$ \cite{Muller, Takahashi2012,Takahashi2008, Kontani2009, Takahashi2002}. Similarly, the complete suppression of spin pumping in $\alpha$ in Fig. \ref{Fig: series4}(a) also first manifests below $T\leq2.5 $ K. The manifestation of the QMiSHE only for a direct contact between SC and FM is in agreement with previous observations in Ref. \onlinecite{Muller}. In comparison to our results for NbN/Py-bilayers in Ref. \onlinecite{Muller}, we find that the QMiSHE is more pronounced in sample A ($\sigma_\mathrm{d}= -(1.2\pm0.1)\times10^5/\;\Omega$m at $T/T_\mathrm{c}\approx0.5$), which we attribute to a larger spin Hall angle in TaN than in NbN due to SOI. Its experimental detection for TaN/Py-bilayers also substantiates the capability of our inductive analysis method to detect the QMiSHE.\\ In Fig. \ref{Fig: series4}(d), we plot the extracted field-like $\sigma_{\mathrm{f}}$ normalized by $\sigma_{\mathrm{300K}}$. We observe small positive values in the normal state that are compatible with the iREE \cite{Bychkov1984a, Edelstein1990}, as the Faraday effect would give rise to negative $\sigma_{\mathrm{f}}$ \cite{Berger2018}. Similar $\sigma_{\mathrm{f}}$ values have already been detected in Pt/Py-bilayers \cite{Berger2018}. In the SC state, we observe a large positive $\sigma_{\mathrm{f}}$. The observation of a field-like $\sigma_{\mathrm{f}}$ in the SC state is in agreement with our previous results on NbN/Py-heterostructures \cite{Muller}. We note that the steep, step-wise increase in $\sigma_{\mathrm{f}}$ below 2.5 K provides further evidence for the presence of a second superconducting transition for $T\leq2.5 $ K, though this effect does not require direct SC/FM-contact \cite{Muller}. From this result, we infer that the magnitude of $\sigma_{\mathrm{f}}$ depends on the superconducting condensate density $n_{\mathrm{s}}$. The derived magnitude for $\sigma_{\mathrm{f}}$ in TaN/Py-bilayers ($\sigma_\mathrm{f}= +(1.1\pm0.1)\times10^5/\;\Omega$m at $T/T_\mathrm{c}\approx0.5$) is lower than that of NbN/Py-bilayers at comparable reduced temperatures $T/T_{\mathrm{c}}$. The origin of the positive $\sigma_{\mathrm{f}}$ in the SC state is as of today unexplained. Potential sources may be the coherent motion of vortices in an rf-field \cite{Dobrovolskiy2018, Awad2011}, the impact of Mei{\ss}ner screening currents on the magnetization dynamics due to triplet superconductivity \cite{Jeon2019b} or non equilibrium effects \cite{Ouassou2020}. Finally we note that abrupt changes in our $\sigma^{\mathrm{SOT}}$ as function of $T$ enable the indirect detection of additional fractional superconducting transitions in SC/FM-heterostructures and thereby to study the SC/FM-proximity effect and thin film homogeneity. \\
\section{Summary}
In summary, by optimizing the reactive dc magnetron sputtering deposition for superconducting TaN films on \ch{SiO2}, we achieved films with a superconducting $T_{\mathrm{c}}$ of up to 5 K and a critical field of $\mu_0 H_{\mathrm{c2}}(T=0)=(13.8\pm0.1)$ T. Performing broadband ferromagnetic resonance experiments as a function of temperature showed that in the superconducting state $\alpha$ decreases due to the freeze-out of thermally excited quasiparticles and a finite QMiSHE manifests. Overall the results of bbFMR spectroscopy are in good agreement to those in our previous study on NbN/Py-bilayers \cite{Muller} and highlight the universality of the observed effects for hard type II s-wave superconductors. Our study demonstrates that TaN is a promising high spin-orbit coupling material for superconducting spintronics. This paves the way towards the study of a manifold of recently proposed exotic phenomena at the SC/FM-interface such as the generation of supercurrents by Rashba spin-orbit interaction \cite{Bychkov1984a, He2019}, supercurrent-induced spin-orbit torques \cite{Hals2016} or the vortex spin Hall effect \cite{Vargunin2019}. 
%\begin{itemize}
%	\item Method by Berger \cite{Berger2018}
%\end{itemize}
%\begin{align}
%\begin{aligned}
%\tilde{L}(f)&=\tilde{L}_0+\tilde{L}_{\mathrm{j}}(f)\\
%&=\tilde{L}_0+f\cdot(\mathrm{Re}(\Delta\tilde{L}_{\mathrm{j}})\epsilon_{\mathrm{r}}(f)+i\cdot\mathrm{Im}(\Delta\tilde{L}_{\mathrm{j}}) \epsilon_{\mathrm{i}}(f)).
%\label{eq: L-tilde}
%\end{aligned}
%\end{align}
%\begin{equation}
%\tilde{L}_\mathrm{j}(f)=C\cdot f\cdot[-\epsilon_{\mathrm{r}}(f)\sigma_{\mathrm{f}}+i\epsilon_{\mathrm{i}}(f)\sigma_{\mathrm{d}}].
%\label{eq: Lj}
%\end{equation}
\nocite{Jeon2019a,Golovchanskiy2020a}
\section*{Supplementary Material}
See the supplementary material for additional details on the growth parameters, supporting bbFMR data and details on the inductive analysis method
\begin{acknowledgments}
We acknowledge financial support by the Deutsche
Forschungsgemeinschaft (DFG, German Research Foundation) via WE5386/4-1 and Germany’s Excellence Strategy EXC-2111-390814868.	
\end{acknowledgments}
\section*{Data Availability}
The data that support the findings of this study are available
from the corresponding author upon reasonable request.
%\section*{Author contributions}

%T.W., M.A., L.L. and N.V. carried out the experiments and evaluated the data. T.W. fabricated the sample. S.G. performed the thin film growth. H.H., M.A. and R.G. supervised the project. T.W., M.A., M.W., R.G., H.H. and S.G. interpreted the results. T.W., M.A., M.W. and H.H. prepared the manuscript and figures with the assistance of all authors.

\bibliography{library}

\end{document}